\documentclass[aip,jmp,showpacs,showkeys,amsmath,preprint,nofootinbib,onecolumn]{revtex4-1}

\begin{document}
   \title{Laplace-Runge-Lenz symmetry in general rotationally symmetric systems}

  \author{Uri \surname{Ben-Ya'acov}}

 \affiliation{School of Engineering, Kinneret Academic College on
   the Sea of Galilee,\\D.N. Emek Ha'Yarden 15132, Israel}

   \email{uriby@tx.technion.ac.il}

    \date{\today}

\begin{abstract}
{The universality of the Laplace-Runge-Lenz symmetry in all rotationally symmetric systems is discussed. The independence of the symmetry on the type of interaction is proven using only the most generic properties of the Poisson brackets. Generalized Laplace-Runge-Lenz vectors are definable to be constant (not only piece-wise conserved) for all cases, including systems with open orbits. Applications are included for relativistic Coulomb systems and electromagnetic/gravitational systems in the post-Newtonian approximation. The evidence for the relativistic origin of the symmetry are extended to all centrally symmetric systems.}
\end{abstract}



To be published in J. Math. Phys.

\maketitle

\section{Introduction}

The symmetry associated with the so-called Laplace-Runge-Lenz
(LRL) vector in Newtonian Kepler-Coulomb 2-body systems \cite{Goldstein02},
well-known for more than two centuries \cite{Goldstein75,Goldstein76}, is regarded by many as 'accidental' or 'hidden'\cite{DulMcInt66,StehleHan67,McIntosh71,Heintz74,KhachidzeKhelash08}. It has gained this adjective being associated originally only with $1/r$ potentials and because of the particular properties of bound states in these interactions -- closed orbits and extra degeneracy of the energy states; and, also, because it had no apparent more profound source than its mere existence, in strong contrast with the common space-time Galilean or Lorentz-Poincar\'{e} symmetries, which are clearly geometrical in
nature.

The realization, following Bacry et al \cite{BacryRS66}, Fradkin \cite{Fradkin67} and Mukunda \cite{Mukunda67a}, that LRL-like vectors do exist and may be defined for \textit{all} systems with rotational symmetry, changed this picture drastically. The generalized LRL vectors generate (together with the internal angular momentum) $o(4)$ or $o(3,1)$ Lie-Poisson algebras just as in the Kepler-Coulomb case, extending the LRL symmetry into various systems, including, in particular, centrally symmetric ones with open orbits.

These developments dissociate the two aspects which in the past seemed, following the Kepler-Coulomb case, to be inseparably characterizing the LRL symmetry -- closed orbits and extra degeneracy on the one hand, and the existence of a constant LRL vector and the algebraic structure associated with it on the other hand. While the extra degeneracy continues to be peculiar to some interactions only, a LRL vector always accompanies internal rotational symmetry.

Following the three seminal papers, generalized LRL symmetry was realized many times in the literature in various systems, including relativistic ones (see, \textit{e.g.}, Refs.~\onlinecite{ArgSanz84,DVNur90,Horwitz93,Duviryak96}), the MICZ system (electric charge + magnetic monopole)~\cite{Zwanziger68,MIC70,Meng07}, systems with spatial constant curvature \cite{IwaiKatayama95,KeaneBarSimm00}, or other systems with velocity and/or time-dependent interactions besides central potentials \cite{LeachFlessas03} (see also Ref.~\onlinecite{KhachidzeKhelash08} for a more exhaustive list).
Once the association between the LRL symmetry and a particular type of interaction has been dissolved, the symmetry may not be regarded any more as 'dynamical' and 'accidental'. Being an inseparable aspect of any rotationally symmetric system, the LRL symmetry must be of geometrical origin and nature.

The first step towards identifying and understanding the physical origin of the LRL symmetry was done by Dahl, who discovered \cite{Dahl97,Dahl68} some years ago for the classical (Kepler-Coulomb) LRL symmetry that its origin resides within the relativistic framework : Dahl has shown that the Newtonian LRL vector appears naturally in the computation of the Lorentz boost in the post-Newtonian approximation of electromagnetic or gravitational 2-body systems. Although it is only the Newtonian LRL vector that appears
there this an essentially relativistic result, because it is of
order $1/c^2$, vanishing in the full non-relativistic limit when
the Lorentz boost becomes the Galilei boost. In essence, Dahl's result stems from the observation that the Newtonian centre-of-mass (CM) of an $N$-body system $\vec X_{\rm N} = \left(\sum_a m_a \vec x_a\right) / \left(\sum_a m_a \right)$ in the CM reference frame, which is not constant for relativistic systems, nevertheless its time varying part must be purely relativistic because $\vec X_{\rm N}$ is constant in the non-relativistic limit. Explicitly computing this time-varying part of $\vec X_{\rm N}$, two independent solutions ensue, the difference of which is proportional to the LRL vector.

The generalizability of the LRL symmetry \textit{a-la} Bacry et al, Fradkin and Mukunda, on the one hand, and the realization of its relativistic origin by Dahl on the other hand, now call for unification. The purpose of the present paper is therefore to demonstrate, confirm and further establish the generality and universality of the Laplace-Runge-Lenz symmetry, bringing together the two approaches.

We start (Sec.\ref{sec: genLRL}) by introducing two propositions which encompass the essence of the LRL symmetry in general rotationally symmetric systems. These propositions provide a novel and very simple demonstration of the generality of the LRL symmetry together with a straight-forward tool to compute the Lie-Poisson brackets of the LRL vector in any rotationally-symmetric constellation. The generic properties and the consequences of the LRL symmetry are discussed in Sec.\ref{sec: algeprop} together with the effect of the corresponding symmetry transformation.

The explicit construction of the LRL vector for general centrally-symmetric 2-body systems starts to be discusses in Sec.\ref{sec: expcon}. The method for constructing the LRL vectors for centrally-symmetric systems with open orbits suggested in the past \cite{Fradkin67,Peres79} deemed these vectors to be only piece-wise conserved, changing directions non-continuously in turning points \cite{StehleHan67,BuchDenman75,Peres79,Yoshida87b,HolasMarch90}, thus reducing substantially any interest in their applicability and usefulness. In fact, almost all the known applications of generalized LRL vectors are in systems for which the vector is constant, corresponding to closed orbits. Since the LRL vectors that are obtained via Dahl's (relativistic) procedure (see Ref.\onlinecite{Dahl97} and Sec.\ref{sec: CMPN} below) are constant, while even for simple relativistic systems the orbits are open\cite{LLfields75,Boyer04}, this situation is not satisfactory. To remedy the situation, we provide a definition for constant LRL vectors in general centrally symmetric systems that is valid also for open orbits, together with the explicit computation of its self PB. This method is then applied in Sec.\ref{sec: Coul} to two relatively simple relativistic systems with open orbits, the relativistic Coulomb system and post-Newtonian electromagnetic or gravitational system. In both cases the LRL vector is explicitly constructed, via two alternative ways.

Finally, we refer in Sec.\ref{sec: CMPN} to the relativistic origin of the LRL symmetry in general centrally symmetric systems, showing that the Newtonian LRL vector found in the preceding sections is indeed derivable from the Lorentz boost of the corresponding post-Newtonian extensions. The article then concludes with a discussion and summation of the generic properties of the LRL symmetry.

\vskip 30pt

\section{\label{sec: genLRL}Generalization of the Laplace-Runge-Lenz symmetry }

To elucidate now the way the LRL symmetry appears, in a natural way, in all rotationally symmetric systems, and to produce a tool which will greatly simplify the discussion of the symmetry in general systems, let us introduce two propositions which sum up the main properties of the LRL symmetry, generalizing the properties of the symmetry in the classical Kepler-Coulomb case.

Rotational symmetry is assumed to be characterized by the existence of constant internal angular-momentum vector $\vec \ell$ (the term \textit{internal} refers here and in the following to any dynamical quantity which depends only on the relative coordinates of the
particles and their relative motion and is invariant under uniform
global translations; in the case of a single particle with a fixed centre of force, all internal quantities are defined relative to the centre of force.). It should be noted that for non-centrally symmetric interactions, say with spin or velocity dependence, $\vec \ell$ is not just the orbital angular momentum with terms like $\vec r \times \vec p$ but includes also extra terms (as in Eq.(\ref{eq: E + ell gen MICZ}) below).

The systems under consideration are also assumed to be endowed with Lie-Poisson brackets (PB) $\{.,.\}$. For the general discussion in the present Section and Sec.~\ref{sec: algeprop} it suffices that these PB satisfy the general requirements from Lie-Poisson brackets \cite{MarsdenRatiu}, and no canonical-simplectic structure needs to be assumed (such a structure, with the standard definition for the PB, will only apply to particular examples later on).

The identification of $\vec\ell$ as the generator of internal spatial rotations, relative to the centre of mass or centre of force, is then incorporated in the requirement for the existence of rotational PB
 \begin{eqnarray} \label{eq: PBKell}
&& \left\{ \ell^i ,\ell^j \right\} = \varepsilon^{ijk} \ell^k \nonumber \\
&& \left\{ K^i ,\ell^j \right\} = \left\{ \ell^i ,K^j
\right\} = \varepsilon^{ijk} K^k \, ,
 \end{eqnarray}
for any internal vector $\vec K$. The anti-symmetry of the self PB $\left\{ K^i,K^j \right\}$ implies, for any internal vector $\vec K$, the existence of another internal vector $\vec \Lambda$ such that
\begin{equation}\label{eq: KK to Lambda}
 \left\{ K^i,K^j \right\} = \varepsilon^{ijk} \Lambda^k
\end{equation}
From Eq.(\ref{eq: KK to Lambda}) it follows, using the vector property
(\ref{eq: PBKell}) for $\vec K$, that
\begin{equation}\label{eq: PB K-ellK}
\left\{ K^i, \vec\ell\cdot \vec K \right\} = \left\{ K^i ,\ell^j
\right\} K^j + \left\{ K^i,K^j \right\} \ell^j = \varepsilon^{ijk}
\Lambda^k \ell^j
\end{equation}
The product $\vec\ell\cdot \vec K$ is therefore $\vec
K$-invariant, in the sense that $\left\{ K^i,\vec\ell\cdot \vec K
\right\} = 0$, iff the vectors $\vec \Lambda$ and $\vec\ell$ are
parallel, say as $\vec \Lambda = \alpha \vec \ell$. We may therefore
conclude that

\vskip10pt \noindent \textbf{ proposition 1} \,
\textit{The self PB of an internal vector $\vec K$ are of
the form}
\begin{equation}\label{eq: PB KK alpha}
 \left\{ K^i,K^j \right\} = \alpha \varepsilon^{ijk} \ell^k
\end{equation}
\textit{with $\alpha$ some scalar, iff the product $\vec\ell\cdot
\vec K$ is $\vec K$-invariant, $\left\{ K^i,\vec\ell\cdot \vec K
\right\} = 0$}.

In 2-body systems any scalar
observable constant of the motion must be functionally dependent
on $H$ and $\vec \ell^2$. Hence, if $\vec K^2$ is a constant of the
motion then it must be a function of $H$ and $\vec
\ell^2$, say $\vec K^2 = F\left(H,\ell^2\right)$. For more general systems, constant scalar observables need not be functionally dependent on $H$ and $\ell^2$ only. However, we may still  consider those vectors $\vec K$ for which the product $\vec \ell\cdot\vec K$ is $\vec K$-invariant, $\{ \vec K ,\vec \ell\cdot \vec K \} = 0$, and their magnitude $\vec K^2$
depends, besides  $H$ and $\ell^2$, also upon other observables which are $\vec \ell-$ and $\vec K-$invariant. These observables will henceforth be generically denoted as $\mathcal{A}$ (thus $\left\{\vec\ell,\mathcal{A}\right\} = 0$, $\left\{\vec K,\mathcal{A}\right\} = 0$), so that we may write now $\vec K^2 = F(H,\ell^2,\mathcal{A})$. Then we have
\vskip1pt \noindent \textbf{ proposition 2} \,
\textit{Let $\vec K$ be an internal vector such that :
\begin{enumerate}
 \item
The product $\vec \ell\cdot\vec K$ is $\vec K$-invariant
 \item
$\vec K^2 = F(H,\ell^2,\mathcal{A})$
\end{enumerate}
Then the self PB of $\vec K$ satisfies}
 \begin{equation}\label{eq: PB KK}
\left\{K^i,K^j\right\} = - \frac{\partial \left(\vec K^2 \right)}{\partial(\ell^2)} \varepsilon^{ijk} \ell^k
 \end{equation}

\noindent \textbf{ Proof} \,
The PB $\left\{ K^i,\vec K^2 \right\}$ may be computed either as
 \begin{equation} \label{eq: PB K K2 1}
\left\{ K^i,\vec K^2 \right\} = 2\left\{ K^i,K^j \right\}K^j = 2\alpha \varepsilon^{ijk} K^j \ell^k
 \end{equation}
or as
 \begin{equation} \label{eq: PB K K2 2}
\left\{ K^i,\vec K^2 \right\} = \left\{
K^i,F\left(H,\ell^2,\mathcal{A}\right) \right\} = \frac{\partial F}{\partial(\ell^2)} \left\{ K^i ,\ell^2 \right\} = 2 \frac{\partial F}{\partial(\ell^2)} \varepsilon^{ijk} \ell^j K^k
 \end{equation}
Comparing the last two equations and taking into account the
non-parallelism of $\vec \ell$ and $\vec K$, Eq.(\ref{eq: PB KK}) follows. QED

\vskip10pt

To illustrate the applicability of these propositions, here are few examples :

\begin{enumerate}
\item{
First, for Newtonian 2-body Kepler-Coulomb systems with the Hamiltonian
 \begin{equation}\label{eq: hamiltonian KC}
H = \frac{p^2}{2\mu} + \frac{\kappa}{r}
\end{equation}
($\vec r = \vec x_1 - \vec x_2$ is the relative coordinate, $\vec p$ the corresponding momentum and $\mu$ the Newtonian reduced mass) and internal angular momentum $\vec\ell = \vec r \times \vec p$, the LRL vector is commonly defined as
 \begin{equation} \label{eq: KC LRL}
\vec K = \vec p \times \vec \ell + \frac{\mu \kappa}{r}\vec r
 \end{equation}
Substituting the magnitude $\vec K^2 = 2\mu H \ell^2 + \mu^2 \kappa^2$ in Eq.(\ref{eq: PB KK}) then yields the well-known result

 \begin{equation} \label{eq: PB KC R-L}
 \left\{K^i,K^j\right\} = - 2 \mu H \varepsilon^{ijk} \ell^k
 \end{equation}
}
\item{
In systems with a magnetic monopole and general central interaction the equation of motion is
 \begin{equation}\label{eq: gen MICZ}
m\frac{d\vec v}{dt} = \frac{\alpha}{r^3}\vec v \times \vec r - \frac{1}{r}U'\left( r \right) \vec r
 \end{equation}
with conserved energy and angular momentum
 \begin{equation}\label{eq: E + ell gen MICZ}
E = \frac{1}{2} m v^2 + U\left( r \right) \qquad , \qquad
\vec \ell  \equiv m\vec r \times \vec v - \frac{\alpha }{r}\vec r
 \end{equation}
In particular, for a MICZ \cite{Zwanziger68,MIC70} system with modified Coulomb interaction
 \begin{equation}\label{eq: U MICZ}
U\left( r \right) = \frac{\kappa}{r} + \frac{\alpha^2}{2m r^2}
 \end{equation}
the LRL vector takes the particularly simple form
 \begin{equation}\label{eq: K MICZ}
\vec K = m \vec v \times \vec \ell  + \frac{m \kappa}{r}\vec r
 \end{equation}
Here $\vec K \cdot \vec \ell = - m \alpha \kappa$ : non-zero, but being composed only of constants of the system it is $\vec K$-invariant. Substituting the magnitude $\vec K^2 = 2mE (\ell^2 - \alpha^2) + m^2 \kappa^2$ in Eq.(\ref{eq: PB KK}) then yields immediately, without any tedious computations,
 \begin{equation} \label{eq: PB MICZ R-L}
 \left\{K^i,K^j\right\} = - 2 m E \varepsilon^{ijk} \ell^k
 \end{equation}
}
\end{enumerate}

In conclusion, we note that :

\begin{enumerate}
\item{
As is evident from its derivation, Eq.(\ref{eq: PB KK}) is independent of any particular form of the interaction, except for the requirement for rotational symmetry.
}
\item{
Eq.(\ref{eq: PB KK}) provides a simple and straight-forward tool to compute the self PB of $\vec K$. Its usefulness will become evident in the various applications in the following. In particular, comparing with Fradkin's\cite{Fradkin67} computations for LRL vectors in general centrally symmetric systems, Eq.(\ref{eq: PB KK}) will prove to save need to launch on very tedious calculations.
}
\item{
Moreover, Eq.(\ref{eq: PB KK}) doesn't require any particular recipe for the computation of the PB. It is therefore also suitable for systems which lack canonical or phase-space structure, such as in some relativistic action-at-a-distance systems.
}
\end{enumerate}

Any internal vector observable $\vec K$ that satisfies these conditions may be regarded as a \textit{generalized Laplace-Runge-Lenz vector}. We now proceed for further applications of these propositions, to establish and confirm the universality of the LRL symmetry.

\vskip30pt

\section{\label{sec: algeprop}Algebraic properties and consequences of the generalized Laplace-Runge-Lenz symmetry}

From the discussion of Sec.~\ref{sec: genLRL} it follows that any constant vector which is not parallel to $\vec\ell$ could serve, at least in principle, as a LRL vector. This becomes evident in the following way: Let $\vec u_o$ be a constant unit vector perpendicular to $\vec\ell$. By Eq.(\ref{eq: PB KK}), any vector whose square is independent of $\ell$ has
vanishing self PB. Therefore, any such constant unit vector is a
generalized LRL vector with vanishing self PB,
 \begin{equation}\label{eq: unit vec}
\vec u_o \cdot \vec \ell = 0  \quad , \quad  \vec u_o \cdot \vec u_o = 1
\qquad \Rightarrow  \qquad  \{u_o^i,u_o^j\} = 0 \, ,
 \end{equation}
and may be used to generate arbitrary generalized LRL vectors via the relation
  \begin{equation}\label{eq: unit to K}
\vec K = f(E,\ell^2,\mathcal{A}) \vec u_o + g(E,\ell^2,\mathcal{A})
\vec\ell \times \vec u_o + \frac{h(E,\mathcal{A})}{\ell^2} \vec \ell
  \end{equation}
The coefficients $f(E,\ell^2,\mathcal{A})$,
$g(E,\ell^2,\mathcal{A})$ and $h(E,\mathcal{A})$ may be arbitrary
functions of their arguments, where $\mathcal{A}$ stands for possible constant scalar observables which are $\vec \ell$- and $\vec K$-invariant. $\vec K \cdot \vec\ell =
h(E,{\mathcal A})$ cannot depend on $\vec\ell^2$ since it must be $\vec K$-invariant.

The LRL vectors constructed in Eq.(\ref{eq: unit to K}) are very general, and in principle, with an appropriate choice of the coefficients one may get almost any desired result. However, the physical meaning of such a construction would be
obscure. We therefore seek now to limit this generality by
introducing some physical content.

By properly choosing the coefficients in Eq.(\ref{eq: unit to K}) and taking Eq.(\ref{eq: PB KK}) into account, a LRL vector $\vec A$ with self PB
 \begin{equation}\label{eq: PBAA}
\left\{A^i,A^j\right\} = \eta \varepsilon^{ijk} \ell^k = \eta
\ell^{ij}
 \end{equation}
may always be created with the coefficient $\eta$ being either $+1$ or $-1$ (the critical limiting value $\eta = 0$ needs not be considered separately). Adding the
PB of rotation,
 \begin{eqnarray} \label{eq: PBAell}
&& \left\{ \ell^i ,\ell^j \right\} = \varepsilon^{ijk}
\ell^k \nonumber\\
&& \left\{ A^i ,\ell^j \right\} = \left\{ \ell^i ,A^j
\right\} = \varepsilon^{ijk} A^k \, ,
 \end{eqnarray}
the vectors $\vec A$ and $\vec\ell$ generate together
classical Lie-Poisson algebras $o(4)$ or $o(3,1)$ according to
wether $\eta = +1$ or $-1$, respectively. Any vector $\vec A$
satisfying Eq.(\ref{eq: PBAA}) may be regarded as a \textit {canonical
Laplace-Runge-Lenz vector}.

In quantum systems, the Casimir operators determine the quantum state of the system. In classical systems, the Casimir invariants of the corresponding Lie-Poisson algebra provide information regarding the physical state of the system. With the PB (\ref{eq: PBAA}) and (\ref{eq: PBAell}), the two Casimir invariants are
\begin{subequations}
 \begin{eqnarray}
&& \mathcal{C}_1 = \mathcal{C}_1 (H,\mathcal{A}) = \eta \vec A\,^2 + \vec \ell^2  \label{eq: Casimir1} \\
&& \mathcal{C}_2 = \mathcal{C}_2 (H,\mathcal{A}) = \vec A \cdot \vec \ell   \label{eq: Casimir2}
 \end{eqnarray}
\end{subequations}
For many 2-body systems, in particular centrally-symmetric ones, the motion is in a plane perpendicular to $\vec \ell$. Then the LRL vectors, and in particular the canonical one $\vec A$, may be chosen in the plane of motion so that $\mathcal{C}_2 = \vec A \cdot \vec \ell = 0$. Other systems, such as the MICZ \cite{Zwanziger68,MIC70} systems discussed in example B in Sec.~\ref{sec: genLRL} [Eqs.~(\ref{eq: gen MICZ}-\ref{eq: K MICZ})] or generalizations thereof such as those discussed by Iwai and Katayama\cite{IwaiKatayama95}, may be characterized by non-zero $\mathcal{C}_2$. However, even in the latter case it follows from Eq.(\ref{eq: unit to K}) that the component of $\vec K$ parallel to $\vec\ell$ may be chosen at will. Therefore, if $\mathcal{C}_2 \ne 0$, we may always use the component of $\vec A$ perpendicular to $\vec\ell$ instead of $\vec A$. In other words, it is always possible to assume $\mathcal{C}_2 = 0$ without loss of generality, and this assumption will be held in the following.

The sign coefficient $\eta$ is not arbitrary, but reflects the energetical state of the system as for classical Kepler-Coulomb systems. For a given value of the total energy $E$, a system may be either bound or un-bound. The $o(4)$ case ($\eta = +1$) corresponds to
bound states : From Eq.(\ref{eq: Casimir1}) it follows that
$\mathcal{C}_1$ is necessarily non-negative, with $\ell^2$ bounded
from above,
 \begin{equation}\label{eq: ell2bound}
    \ell^2 \le \mathcal{C}_1
 \end{equation}
The existence of an upper limit for $\ell$ (for a given value of $E$) is characteristic of
bound states. In particular, in the case of circular motion rotational invariance requires $\vec A = 0$. Then, with given total energy $E$, $\ell$ achieves its maximal value\cite{Bacry91}. Denoting this maximal value as $\ell_{\text{max}}(E)$ it determines $\mathcal{C}_1$ as\cite{Fradkin fn}
 \begin{equation}\label{eq: C1E}
   \mathcal{C}_1 (E) = \ell_{\text{max}}^2(E)
 \end{equation}
The opposite case ($\eta = -1$), with the $o(3,1)$ algebra,
corresponds to un-bound states: Since $\vec A^2 = \ell^2 -
\mathcal{C}_1 \ge 0$ there is no limitation on $\ell$, but
$\mathcal{C}_1$ must be negative in order to allow situations like
head-on collisions in which $\ell = 0$. Still, except for the
sign, the functional dependence of $\mathcal{C}_1$ on $E$ is the
same as in Eq.(\ref{eq: C1E}).

Another aspect of the general construction of the LRL vector in Eq.(\ref{eq: unit to K}) is that its direction may also be chosen at will. It is evident that the PB of $\vec A$ in Eqs.~(\ref{eq: PBAA}) and (\ref{eq: PBAell}) do not change if $\vec A$ is rotated perpendicular to $\vec\ell$. Therefore, at least as far as the algebraic relations are concerned, the LRL symmetry does not distinguish any particular direction for the LRL vector.

There is, however, a way to determine a preferred direction for the LRL vector : From the post-Newtonian derivation by Dahl \cite{Dahl97} emerges the classical LRL vector (\ref{eq: KC LRL}), directed towards the perihelion. Later, in Sec.\ref{sec: CMPN}, it is verified that the LRL vector that emerges from a similar derivation for general centrally-symmetric systems also points in the direction of closest approach ('generalized perihelia').

We conclude the present section with a discussion of the effect of the LRL symmetry on its own algebra. Symmetries in classical dynamics were firstly manifested in terms of mappings of the configuration- or phase-space. This has certainly been the case with the common space-time symmetries (translations, rotations, dilatations, etc.). In contrast, with the LRL symmetry the focus has always been on the constant LRL vector and the corresponding algebraic structure that it generates together with the internal angular momentum. The transformations which constitute the LRL symmetry group for classical Kepler-Coulomb systems were discussed \cite{LevyLeblond71,PrinceEliezer81,Krause94} only many years after the discovery of the algebra that dominates the symmetry.

In the foregoing discussion we considered the generic properties of the LRL symmetry, so generic even to the extent of not assuming any particular structure for the configuration- or phase-space. But even in the absence of such structure, the algebra can provide an insight into the nature of these transformations. While $\vec \ell$ generates internal rotations, the internal transformations that
correspond to the LRL symmetry are generated by $\vec A$. Let $\vec n$ be
some constant unit vector and $\chi$ be a real dimensionless
parameter. The infinitesimal transformations generated by $\vec A$
relative to the direction of $\vec n$ via the
generator $\delta {\mathcal G} = \vec n \cdot \vec A \delta\chi$
satisfy the equations
 \begin{equation}  \label{eq: ell & A trans}
 \delta \vec \ell = \left\{\vec \ell, \delta {\mathcal G}\right\} =
 \vec n \times \vec A \delta \chi \qquad , \qquad  \delta \vec A = \left\{\vec A, \delta {\mathcal G}\right\} =
 \eta \vec n \times \vec \ell \delta \chi
 \end{equation}
The orbits of these transformations within the algebra, in terms of $\vec A$ and $\vec\ell$, are given by Rodrigues-like formulae
\begin{subequations}\label{eq: LRL gen trans}
 \begin{eqnarray}
 \eta = +1 \, :  \nonumber \\
& \vec \ell (\chi) = \cos \chi \cdot \vec \ell (\chi = 0) + (1 -
 \cos \chi) \left[\vec \ell (\chi = 0) \cdot \vec n \right] \vec n + \sin
 \chi \vec n \times \vec A (\chi = 0) \nonumber  \\
& \vec A (\chi) = \cos \chi \cdot \vec A (\chi = 0) + (1 -
 \cos \chi) \left[\vec A (\chi = 0) \cdot \vec n \right] \vec n + \sin
 \chi \vec n \times \vec \ell (\chi = 0) \label{eq: A+el bound} \\
  \eta = -1 \, :  \nonumber \\
& \vec \ell (\chi) = \cosh \chi \cdot \vec \ell (\chi = 0) - (\cosh
\chi - 1) \left[\vec \ell (\chi = 0) \cdot \vec n \right] \vec
n + \sinh \chi \vec n \times \vec A (\chi = 0) \nonumber \\
& \vec A (\chi) = \cosh \chi \cdot \vec A (\chi = 0) - (\cosh
 \chi - 1) \left[\vec A (\chi = 0) \cdot \vec n \right] \vec n
 - \sinh \chi \vec n \times \vec \ell (\chi = 0) \label{eq: A+el unbound}
 \end{eqnarray}
\end{subequations}
These transformations, relating orbits with equal energy but different angular momentum, may be regarded as deforming transformations. They are the non-quantum analog of the shift operators generated by the LRL vector for the hydrogen atom wave function  \cite{Stahlhofen1998,LiuLeiZeng1998,BurkLeven2004,Stahlhofen05b}, extending to arbitrary centrally symmetric potentials.

$\vec A$-generated transformations may change the direction of $\vec\ell$. For centrally-symmetric 2-body systems this implies changing the orientation of of the plane of motion, which is merely a geometrical transformation rather than physical. In such systems we may always choose the vectors $\vec A$ and $\vec n$ perpendicular to $\vec\ell$ thus restricting the algebra, without loss of generality, to the subalgebra defined by $\mathcal{C}_2 = \vec A \cdot \vec \ell = 0$. The vector $\vec \ell$ may then change its magnitude but not the direction, keeping the plane of motion intact. The physical system is then fully characterized by the first Casimir invariant $\mathcal{C}_1$, while different states or configurations are characterized by $\vec \ell$ and $\vec A$.

In the case of bound centrally symmetric systems it is convenient to start, at $\chi =
0$, at the state of circular motion for which $\vec A = 0$. We may
be interested in particular to maintain the direction of the
angular momentum unchanged, which implies that $\vec n$ must be
perpendicular to $\vec \ell$. Then, using $\vec \ell (\chi = 0) =
\ell_{\text{max}}(E) \hat \ell$, we obtain from Eq.(\ref{eq: A+el bound})
 \begin{eqnarray} \label{eq: A+el bound from circ}
& \vec \ell (\chi) & = \ell_{\text{max}}(E) \cos \chi \cdot \hat \ell \nonumber \\
& \vec A (\chi) & = \ell_{\text{max}}(E) \sin \chi \vec n \times \hat \ell =
 \tan \chi \vec n \times \vec \ell = \sqrt{\ell_{\text{max}}^2(E) - \ell^2} \, \vec
 n \times \hat \ell
 \end{eqnarray}

\vskip30pt

\section{\label{sec: expcon}Explicit construction and computation of generalized Laplace-Runge-Lenz vectors}

Fradkin \cite{Fradkin67}, and later also Peres \cite{Peres79} and Yoshida \cite{Yoshida87b}, suggested and discussed a general method for explicitly constructing the generalized LRL vectors for general centrally-symmetric systems with open orbits. This method of construction deemed these vectors to be only 'piece-wise' conserved, changing directions non-continuously in turning points \cite{StehleHan67,BuchDenman75,Peres79,Yoshida87b,HolasMarch90}, thus reducing substantially any interest in their applicability and usefulness. Consequently, almost all the known applications of generalized LRL vectors are in systems for which the vector is constant, corresponding to closed orbits. However, taking into account the fact that the LRL vectors that are obtained via Dahl's (relativistic) procedure (see Ref.\onlinecite{Dahl97} and Sec.\ref{sec: CMPN} below) are constant, together with the fact that even for simple relativistic systems the orbits are open\cite{LLfields75,Boyer04,Stahlhofen05a}, and the results of the Secs.~\ref{sec: genLRL} and \ref{sec: algeprop} implying that constant LRL vectors should exist and be definable for any rotationally symmetric system, this situation calls for a remedy.

The purpose of the present Section is therefore to provide an improved definition of the generalized LRL vector, insuring its constancy for any relevant system. Applying Eq.(\ref{eq: PB KK}), an expression for the self PB of the LRL vector is readily obtained.

In the following we consider 2-body centrally symmetric systems, with
canonical internal variables $\left(\vec r,\vec p\right)$ and conserved
angular momentum $\vec\ell = \vec r \times \vec p$. The construction of LRL vectors in general systems commonly starts \cite{Fradkin67,Peres79,KhachidzeKhelash08} with choosing an arbitrary constant unit vector $\vec u_o$, in any direction perpendicular to $\vec\ell$, as in Eq.(\ref{eq: unit to K}). As we have shown, any such vector satisfies Eq.(\ref{eq: unit vec}) and may be regarded as a LRL vector. With the motion being confined to the plane perpendicular to $\vec\ell$, let $\theta$ be the azimuthal angle defined counter-clockwise from $\vec u_o$ to $\vec r$. In terms of $\vec u_o$ and $\theta$, the unit vector $\hat r = \vec r / r$ may be represented as
 \begin{equation}\label{eq: r theta uo}
 \hat r = \cos \theta \vec u_o + \sin \theta \hat
\ell \times \vec u_o = {\mathcal U}(\theta) \vec u_o
 \end{equation}
where $\hat\ell \equiv \vec\ell / \ell$ and
 \begin{equation}\label{eq: rot op}
 {\mathcal U}(\theta) = \cos \theta + \sin \theta \hat \ell \times
 \end{equation}
is the rotation operator which rotates vectors in the plane of motion counter-clockwise with angle $\theta$. Inverting Eq.(\ref{eq: r theta uo}), it is possible to express $\vec u_o$ in terms of $\hat r$ and $\theta$,
 \begin{equation}\label{eq: unit vec ob}
 \vec u_o(\vec r, \theta) = {\mathcal U}(- \theta) \hat r = \cos \theta \hat r - \sin \theta \hat \ell \times \hat r
 \end{equation}
Taking the constant unit vector $\vec u_o$ as the direction of the
desired generalized LRL vector, and using arbitrarily any
scalar function $K(E,\ell^2)$ for its magnitude, we obtain the
vector
 \begin{equation} \label{eq: vec K general 1}
\vec K = K(E,\ell^2) \cdot \vec u_o = K {\mathcal U}(- \theta) \hat r = K\cos \theta \hat r - K\sin \theta \hat \ell \times \hat r
\end{equation}
which satisfies all the requirements from a generalized LRL
vector. Using the identity
\[ \vec \ell \times \vec r = \frac{\ell^2 \vec r - r^2 \vec p\times \vec \ell}{\vec r \cdot \vec p}
\]
and the notation $p_r = \hat r \cdot \vec p$, it may be brought to a more familiar form
 \begin{equation} \label{eq: vec K general 2}
\vec K = \frac{K \sin \theta}{\ell p_r } \vec p \times \vec \ell +
\left(K\cos \theta - \frac{K\ell \sin \theta}{r p_r} \right) \hat r
\end{equation}
which resembles the classical LRL vector (\ref{eq: KC LRL}).

Though similar to the construction of Fradkin \cite{Fradkin67} and Peres \cite{Peres79} (shown to be equivalent by Yoshida \cite{Yoshida87b}), this construction differs in one essential aspect -- the scalar coefficients of $\vec p \times \vec \ell$ and $\hat r$ in Eq.(\ref{eq: vec K general 2}) are functions of the polar canonical variables $\left(r,\theta;p_r,p_\theta=\ell\right)$, while in the construction of Fradkin and Peres the corresponding coefficients are only functions of $r$, after having integrated the equations of the orbit for the particular orbit. Our construction need not take into account the equations of motion, and it doesn't suffer, therefore, from the drawbacks of the Fradkin-Peres construction -- namely, being only piece-wise conserved. In fact, the construction in Eq.(\ref{eq: vec K general 2}) is just a complicated way to write the constant vector $\vec K = K(E,\ell^2) \cdot \vec u_o$ (\ref{eq: vec K general 1}) which, by virtue of Eq.(\ref{eq: PB KK}), is already recognized as a LRL vector.

So far, the direction of the unit vector $\vec u_o$ in the plane of motion is completely arbitrary. In order to fix it, we now apply the equations of motion.

For an arbitrary central potential the orbit may always be represented as $r = r \left(\theta,E,\ell\right)$. This is a periodic function in $\theta$, with period $\Theta$ being some general function depending on the parameters of the orbit, $\Theta\left(E,\ell\right)$. Expressing $r$ as a function of $\theta$ via $r \left(\theta,E,\ell\right)$, the coefficient of $\vec p \times \vec \ell$ in Eq.(\ref{eq: vec K general 2}) is then in general a varying function of $\theta$, constant only for $1/r$ potentials, as in Eq.(\ref{eq: KC LRL}). Although the vector (\ref{eq: vec K general 2}) is certainly regular, by its construction, the separate coefficients there may be singular because $p_r$ vanishes at the turning points of the orbit. In order to make the coefficients regular we require that $\sin \theta$ also vanish there, which
is achieved by choosing $\vec u_o$ so that it points towards a
turning point. The speciality of the $1/r$ potentials is that only in this case $\Theta = 2\pi$ so $\sin \theta$ may be made to vanish simultaneously at all the turning points; for all other potentials $\sin \theta$ may be made to vanish at one
turning point, but there will be other turning points (if the
system is bound) for which $(\sin \theta) / p_r$ is singular.

To obtain explicit expressions in Eq.(\ref{eq: vec K general 2}) for particular systems, we start from the fact that in any non-circular configuration of the systems in question there is at least one point of minimal approach. For unbound systems
there is just one point like this, which is
the only turning point. For bound systems, except for $1/r$
potentials, there are multiple points of minimal approach, so we
choose arbitrarily one of them. Let the chosen minimal approach
point be at $\vec r = \vec r_m$, directed from the centre-of-mass.

Due to the central symmetry it is convenient to use polar
coordinates, with an Hamiltonian of the most general form, $H
= H\left( r,p_r,p_\theta \right)$ and $p_\theta = \ell$ is a constant since $\partial H/\partial \theta = 0$. Assuming symmetry under spatial reflections and time
reversal, $H$ must be an even function in $p_r$ and $p_\theta$.
Let us choose $\vec u_o = \hat r_m$, the unit vector along $\vec
r_m$, so that $\theta = 0$ there. $p_r$ vanishes at all the turning points since $\dot r = \partial H / \partial p_r$ is odd in $p_r$, and $r_m$ is determined as a function of $E$ and $\ell^2$ as the smallest positive root of the equation
 \begin{equation} \label{eq: rm E el}
 H\left(r_m, p_r=0 , \ell  \right) = E \quad
 \Rightarrow \quad r_m = r_m \left( E,\ell^2 \right)
 \end{equation}
\textit{A-priori}, the magnitude of $\vec K$ may be chosen at will. To fix it with a vector $\vec K$ that resembles the classical LRL vector (\ref{eq: KC LRL}) the most we introduce the condition that the coefficient of $\vec p \times \vec\ell$ in Eq.~(\ref{eq: vec K general 2}) be equal to 1 at $\vec r = {\vec r}_m$,
 \begin{equation}
 \left( \frac{K \sin \theta}{\ell p_r}
 \right)_{r = r_m , \theta = 0} = 1 \, ,
\end{equation}
so that $K$ is determined by
 \begin{equation}  \label{eq: K def gen}
K \left( E,\ell^2 \right) = \ell \left( \frac{dp_r}{d\theta}
\right)_{r = r_m  , \theta = 0}
 \end{equation}
Then, using Hamilton's equations from which follows that
 \begin{equation}  \label{eq: dpr/dtheta gen}
\frac{d p_r}{d\theta} = - \frac{\partial H / \partial r}{\partial H / \partial \ell} \, ,
 \end{equation}
the resultant LRL vector is
 \begin{equation} \label{eq: K vec 1 gen}
\vec K = K \left( E,\ell^2 \right) \hat r_m = - \frac{\ell}{r_m}
\left( \frac{\partial H / \partial r}{\partial H / \partial \ell}
\right)_{r = r_m} \vec r_m
 \end{equation}

These results apply to any centrally-symmetric system. For more specific results, let us consider Newtonian systems with Hamiltonian
 \begin{equation}\label{eq: hamiltonian 2-body}
H = \frac{p^2}{2\mu} + U(r)
\end{equation}
The equation for the minimal distance is
 \begin{equation} \label{eq: pr to rm}
 \left( p_r^2 \right)_{r = r_m} =
2\mu \left[ {E - U(r_m)} \right] - \frac{\ell^2}{{r_m}^2}  = 0
 \end{equation}
so that the magnitude of the generalized LRL vector becomes
 \begin{equation}  \label{eq: K def}
K \left( E,\ell^2 \right) = \frac{ \ell^2}{r_m } - \mu
{r_m}^2 U'\left( r_m \right)
 \end{equation}
and the resultant LRL vector is then
 \begin{eqnarray} \label{eq: K vec 1}
 \vec K && = \left[\frac{\ell^2}{r_m } - \mu {r_m}^2 U'\left( r_m
\right) \right] \left(\cos \theta \hat r - \sin \theta \hat \ell \times \hat r\right) =  \nonumber \\
&& = \left[ \frac{\ell^2}{{r_m}^2 } - \mu r_m U'\left( r_m \right)
\right] \vec r_m
\end{eqnarray}

For $1/r$ potentials the classical LRL vector (\ref{eq: KC LRL}) is evidently obtained in Eq.(\ref{eq: K vec 1}). We also note that for a general potential $U(r)$, in the case of circular motion for which $p_r = 0$ everywhere, $r_m \left[ E,\ell_{\text{max}}^2(E) \right] = r_o$ is the radius of the orbit. Then also $ K $ vanishes identically, $ K \left[ {E,\ell _{\rm max}^2(E)} \right] = 0 $. Otherwise, since $ r \ge r_m $, $ K > 0 $.

The self PB of the vector $\vec {K}$ in Eq.(\ref{eq: K vec 1 gen}) may be
computed by application of Eq.(\ref{eq: PB KK}) and using Eqs.~(\ref{eq: rm
E el}) and (\ref{eq: K def gen}). In particular, for centrally symmetric Newtonian systems we get from Eqs.~(\ref{eq: pr to rm}) and (\ref{eq: K def})
\[
\left. \frac{\partial K }{ \partial \left( \ell^2 \right)} \right]_E
= \frac{\ell^2 - \mu \left[ {r_m}^4 U'\left( r_m \right) \right]'}{2{r_m}^2 K} = \frac{2\mu E - \mu \left[ {r_m}^2 U\left( r_m \right) \right]''}{2 K}
\]
so that the self PB of the vector $\vec K$ (Eq.(\ref{eq: K vec 1})) are
 \begin{equation} \label{eq: gen PBKK}
 \left\{ K^i ,K^j \right\} =  - \frac{\partial \left( K^2 \right)}{\partial \left( \ell^2 \right)} \varepsilon^{ijk} \ell^k = - \left\{ 2\mu E - \mu \left[ r_m^2 U\left( r_m \right) \right]'' \right\} \varepsilon^{ijk} \ell^k
\end{equation}
It is interesting to note that Eq.(\ref{eq: PB KC R-L}) is obtained
not only for $1/r$ potentials, but for all $1/r + 1/r^2$
potentials. This is due to the fact that the $1/r^2$ term causes
the conic section, which is the orbit for the $1/r$ term alone, to
rotate in constant angular rate but leaves the shape of the conic
section intact \cite{Heintz76,Yoshida87a}.

Another way to construct the generalized LRL vectors, which is useful for the post-Newtonian computation in Sec.\ref{sec: CMPN}, starts with the interaction-free part $\vec p \times \vec \ell$ of the classical Kepler-Coulomb LRL vector (\ref{eq: KC LRL}). With the Hamiltonian (\ref{eq: hamiltonian 2-body}) its time derivative is
 \begin{eqnarray} \label{eq: dKkin/dt}
\frac{d}{dt} \left( \vec p \times \vec \ell \right) & = & - \frac{1}{r} U'\left(r\right) \vec r \times \vec \ell = \frac{1}{r} U'\left(r\right) \left[ r^2 \vec p - \left(\vec r \cdot \vec p \right) \vec r \right] = \nonumber \\
& = & \mu r U'\left(r\right) \frac{d \vec r}{dt} - p_r U'\left(r\right) \vec r = \frac{d}{dt} \left[ \mu r U'\left(r\right) \vec r \right] - p_r \left[ r U
\left( r \right) \right]'' \vec r
\end{eqnarray}
Let $\vec W \left(r,\theta,p_r,p_\theta=\ell \right)$ be a vector observable which satisfies
\begin{equation} \label{eq: dW/dt}
\frac{d\vec W}{dt} = p_r \left[ rU \left( r \right) \right]'' \vec r
\end{equation}
with the initial condition $\vec W \left( \theta = 0 \right) = 0$. Then the vector
 \begin{equation} \label{eq: K vec 2}
\vec K = \vec p \times \vec \ell - \mu rU' \left( r \right) \vec r + \vec W = \left[p^2 - \mu rU' \left( r \right)\right] \vec r - \left(\vec p \cdot \vec r\right) \vec p + \vec W
 \end{equation}
is clearly a constant of the motion. For $1/r$ potentials the RHS of Eq.(\ref{eq: dW/dt}) vanishes identically, and the vector in Eq.(\ref{eq: K vec 2}) reduces to the classical LRL vector (\ref{eq: KC LRL}) with $\vec W = 0$. Otherwise, for a general potential $U \left( r \right)$, the two vectors in Eqs.~(\ref{eq: K vec 1}) and (\ref{eq: K vec 2}) coincide at $\vec r = \vec r_m$; and, being both constant, are necessarily identical. This establishes the existence of $\vec W$, and comparing the vectors in Eqs.~(\ref{eq: K vec 2}) and (\ref{eq: vec K general 2}), with $K$ in the latter given by Eq.(\ref{eq: K def}), may be used to obtain an explicit expression for $\vec W$.

It is instructive to verify directly that $\vec W$ is indeed regular in the neighbourhood of $\vec r_m$. Converting the time derivative in Eq.(\ref{eq: dW/dt}) into an angular derivative using the equation of motion for $\theta$,
\[
\frac{d\theta}{dt} = \frac{\ell}{\mu r^2} \, ,
\]
we obtain
\begin{equation} \label{eq: dW/dtheta}
\frac{d\vec W}{d\theta} = \frac{\mu r^2 p_r}{\ell} \left[ rU
\left( r \right) \right]'' \vec r
\end{equation}
$p_r$ vanishes at the turning points, but its angular derivative
 \begin{equation} \label{eq: pr - theta}
\frac{d p_r}{d\theta} = \frac{\ell}{r} - \frac{ \mu r^2 U'\left( r \right)}{\ell} \, ,
 \end{equation}
does not (for non-circular configurations). Then $p_r = O(\theta)$ near $\vec r = \vec r_m$ and consequently $\vec W (\theta) = O\left(\theta^2\right)$ there.

We conclude this section with a recipe for the PB of the generalized LRL vector $\vec K =K \vec u_o$ with any desired observable ${\mathcal F}$. Towards this end, it is most convenient to use Eq.(\ref{eq: unit vec ob}) for the representation of the unit vector $\vec u_o$ as an observable. The computation of derivatives of functions of
$\theta$ is performed using the geometrical relation
 \begin{equation}\label{eq: dtheta}
d\theta = \frac{\left(\hat\ell \times \vec r\right) \cdot d\vec r}{\left(\hat\ell \times \vec r\right)^2}
\end{equation}
and we obtain, after some algebra, for the PB of any observable
${\mathcal F}$ with $\vec u_o$,
 \begin{equation} \label{eq: PB uo gen F}
 \left\{ {\mathcal F},\vec u_o \right\} = - \frac{\sin\theta}{
\ell^3} \left[ \left( r \frac{\partial{\mathcal F}}{\partial \vec
r} + p_r \frac{\partial{\mathcal F}}{
\partial \vec p} \right) \cdot \vec \ell \right] \vec \ell
\end{equation}
Hence, the PB of ${\mathcal F}$ with $\vec K$ given by Eq.(\ref{eq: vec K general 1}) are
 \begin{equation}\label{eq: PB K gen A}
\left\{ {\mathcal F},\vec K \right\} = \left[ \frac{\partial K
}{
\partial H} \left\{ {\mathcal F},H \right\} + \frac{\partial K}{
\partial (\ell^2)} \left\{ {\mathcal F},\ell^2 \right\} \right] \frac{\vec K}{K}
- \frac{K \sin\theta}{\ell^3} \left[ \left( r \frac{\partial{\mathcal
F}}{\partial \vec r} + p_r \frac{\partial{\mathcal F}}{
\partial \vec p} \right) \cdot \vec \ell \right] \vec \ell
\end{equation}

\vskip30pt

\section{\label{sec: Coul}The generalized Laplace-Runge-Lenz symmetry in some exemplary systems}

The LRL symmetry for general Newtonian centrally symmetric systems was discussed in Sec.~\ref{sec: expcon}. The relativistic origin of the LRL symmetry, first realized by Dahl\cite{Dahl97} for Coulomb-Kepler systems and generalized to arbitrary centrally-symmetric systems in the following (Sec.\ref{sec: CMPN}), leads us to consider its appearance in relativistic systems with particular interest. Relativistic systems, even simple ones like the relativistic Coulomb system or the post-Newtonian systems, are characteristically endowed with bound states with open orbits. Still, the LRL vector that emerges from Dahl's procedure is constant. Thus it was important to insure in Sec.~\ref{sec: expcon} that we are equipped with a valid definition for constant LRL vectors.

So far, when the LRL vector was considered in a system with open orbits, the attitude was to start with the classical LRL vector (\ref{eq: KC LRL}) and to follow its rotation\cite{Heintz76,Yoshida87a,Yoshida88b}. The constant LRL vectors computed by the method developed above certainly reduce to the classical LRL vector in the limit of closed orbits, and otherwise, already contain, built in, the data about the rotation of the orbit. We now derive and discuss, as a demonstration and application of the results and methods developed above, the generalized LRL vector in these two relativistic systems.

\vskip20pt

\subsection{\label{subsec: Coul}Laplace-Runge-Lenz symmetry in relativistic Coulomb systems}

Let us consider a relativistic Coulomb system with one of the particles having infinite mass and located at rest at the centre-of-mass. The dynamics of the other particle with mass $m$ are determined by the Hamiltonian (in the present Subsection the convention $c = 1$ is used)
 \begin{equation} \label{eq: H Coul}
 H = \sqrt {p^2 + m^2} + \frac{\kappa}{r} = \sqrt {p_r^2+ \frac{p_\theta^2}{r^2} + m^2} + \frac{\kappa}{r}
 \end{equation}
This is a well-known text-book problem\cite{LLfields75} characterized by irregular orbits, including open orbits for bound states\cite{Boyer04}. So far, only rotating LRL vectors were considered\cite{Yoshida88b,Stahlhofen05a}, attached to the axis of the orbit. In the following we construct the constant LRL vector.

In a configuration with given energy $H = E$ and internal angular momentum $p_\theta =\ell$, the squared radial momentum is isolated as
 \begin{equation} \label{eq: pr2 Coul}
p_r ^2  = \left( E - \frac{\kappa}{r} \right)^2 - m^2 - \frac{\ell^2}{r^2}  = E^2 - m^2 - \frac{2\kappa E}{r} - \frac{\ell^2 - \kappa^2}{r^2} \, ,
 \end{equation}
The type of the orbit depends on the relative values of $E$ and $M$, $\ell$ and $\kappa$. The major relativistic effect is the reduction of the centrifugal barrier from the Newtonian value $\ell^2/r^2$ to $\left(\ell^2 - \kappa^2\right) / r^2$. If $\ell > \left|\kappa\right|$, so the centrifugal barrier is still maintained, the orbits are precessing conic sections. This is a well-known text-book result \cite{LLfields75}. If the centrifugal barrier disappears ($\ell = \left|\kappa\right|$) or even reverses ($\ell < \left|\kappa\right|$, becoming kind of "centrifugal propeller") the orbits become irregular \cite{Boyer04}.

The LRL vector is determined by the turning points, which are determined by the equation $p_r = 0$. In those orbits in which there is a distance of closest approach (otherwise simply $r>0$), it is given by
 \begin{equation} \label{eq: rm Coul}
 r_m = \frac{E\kappa + \sqrt{\left( E^2 - m^2 \right)\ell^2 + m^2 \kappa^2}}{E^2 - m^2}
 \end{equation}
Once the direction of closest approach is fixed (in some arbitrary direction), the LRL vector $\vec K$ is determined by Eq. (\ref{eq: K vec 1 gen}) with the magnitude
 \begin{equation} \label{eq: magK Coul}
 K \left( E,\ell^2 \right) = - \ell \left( \frac{\partial H / \partial r}{\partial H / \partial \ell} \right)_{r = r_m} = \sqrt{\left( E^2 - m^2 \right)\ell^2 + m^2 \kappa^2} \, ,
 \end{equation}
computed from the Hamiltonian~(\ref{eq: H Coul}) and using Eq.(\ref{eq: rm Coul}). Its self PB are then found, applying Eq.(\ref{eq: PB KK}) and using Eq.(\ref{eq: magK Coul}), to be
 \begin{equation}\label{eq: PB KK Coul}
\left\{K^i,K^j\right\} = - \left( E^2 - m^2 \right)\varepsilon^{ijk} \ell^k
 \end{equation}
The type of the algebra is therefore determined solely by the energetic state of the system; although the forms of the orbits depend crucially also on the value of $\ell^2 - \kappa^2$, the algebra is independent of the centrifugal condition. The LRL transformations (\ref{eq: LRL gen trans}) may therefore take the system across the critical point of $\ell = \left|\kappa\right|$  without difficulty. The magnitude of the canonical LRL vector satisfies
 \begin{equation} \label{eq: magA Coul}
 A^2\left( E,\ell^2 \right) = \frac{m^2 \kappa^2}{\left| E^2 - m^2 \right|} - \eta \ell^2 \, ,
 \end{equation}
with coefficient $\eta = - \text{sign} \left( E^2 - m^2 \right)$, in complete agreement with Eq.(\ref{eq: PBAA}), and the first Casimir invariant being
 \begin{equation}\label{eq: C1E Coul}
 \mathcal{C}_1 (E) = \frac{m^2 \kappa^2}{\left| E^2 - m^2 \right|}
 \end{equation}
It is easily verified that for bound states Eq.(\ref{eq: C1E}) is satisfied, since for circular motion
 \begin{equation}\label{eq: ellmax Coul}
 \ell_{\text{max}}(E) = \frac{m \left|\kappa\right|}{\sqrt{m^2 - E^2}}
 \end{equation}
and otherwise $\ell <  \ell_{\text{max}}(E)$. Since $\vec \ell = \vec r \times  \vec p$ and the motion is always in a plane perpendicular to $\vec \ell$, $\vec K \cdot \vec \ell = 0$ and the second Casimir invariant $ \mathcal{C}_2 (E)$ vanishes.

The generic form of the LRL vector is given by Eq.(\ref{eq: vec K general 2}). Since the orbits are open, some dependence of $\vec K$ on the azimuthal angle $\theta$ is unavoidable even when details of the orbit are taken into account. More concrete expressions for the LRL vector depend on the type of orbit. Using the relation
 \begin{equation} \label{eq: pr Coul}
p_r = \frac{\ell}{r^2} \frac{dr}{d\theta}
 \end{equation}
which follows from Hamilton's equations, the equation for the orbit is obtained from Eq.(\ref{eq: pr2 Coul}),
 \begin{equation} \label{eq: geneq Coul}
\left( \frac{1}{r^2} \frac{dr}{d\theta} \right)^2 + \left( 1 - \frac{\kappa^2}{\ell^2}
\right) \frac{1}{r^2} + \frac{2\kappa E}{\ell^2 r} = \frac{E^2 - m^2}{\ell^2}
 \end{equation}
Assuming that $\ell^2 > \kappa^2$, so that a centrifugal barrier does exist, the solution for the orbit is
 \begin{equation} \label{eq: orbit1 Coul}
 \frac{1}{r} +\frac{E\kappa}{\ell^2 - \kappa^2} =  \frac{K \left( E,\ell^2 \right)}{\ell^2 - \kappa^2} \cos \left( \sqrt {1 - \frac{\kappa^2}{\ell^2}} \theta \right)
 \end{equation}
with $K \left( E,\ell^2 \right)$, the magnitude of the LRL vector, given by Eq.(\ref{eq: magK Coul}). Defining the angle
\[
\varphi \equiv \sqrt {1 - \frac{\kappa^2}{\ell^2}} \theta \, ,
\]
the orbit is a conic section which is fixed in the $r-\varphi$ plane, but rotating in the $r-\theta$ plane. To account for the rotation, let $ \psi \equiv \theta - \varphi$. Using the rotation operator identity $\mathcal{U} (\alpha + \beta) = \mathcal{U} (\alpha) \mathcal{U} (\beta)$ and Eq.(\ref{eq: vec K general 1}), the LRL vector may be written as
\[
 \vec K = \mathcal{U}(-\theta) \hat r_m = \mathcal{U}(-\psi - \varphi) \hat r_m = \mathcal{U}(-\psi) \mathcal{U}(-\varphi) \hat r_m
\]
Then, using the relations
\begin{subequations}
\begin{eqnarray}
&& K \cos \varphi = \frac{\ell^2 - \kappa^2}{r} + \kappa E
\\
&& K \sin \varphi = \sqrt{1 - \frac{\kappa^2}{\ell^2}} \frac{\ell^2}{r^2} \frac{dr}{d\theta} = \sqrt{\ell^2 - \kappa^2} p_r
\end{eqnarray}
\end{subequations}
which follow from Eq.(\ref{eq: orbit1 Coul}) together with Eq.(\ref{eq: pr Coul}), the vector
\begin{eqnarray}\label{eq: K' Coul}
\vec {K'} \equiv \mathcal{U}(-\varphi) \hat r_m & = & \frac{K \sin \varphi}{\ell p_r } \vec p \times \vec \ell + \left(K\cos \varphi - \frac{K\ell \sin \varphi}{r p_r} \right) \hat r =  \nonumber \\
& = & \sqrt{1 - \frac{\kappa^2}{\ell^2}} \vec p \times \vec \ell +
\left(\frac{\ell^2 - \kappa^2 - \ell \sqrt{\ell^2 - \kappa^2}}{r} + \kappa E \right) \hat r
\end{eqnarray}
is identified as the constant LRL vector in the fictitious $r-\varphi$ plane, but it rotates together with the conic section in the physical $r-\theta$ plane. $\vec K'$ coincides with the result of Yoshida~\cite{Yoshida88b}, obtained in a much more complicated way. Thus, finally, the LRL vector may be written, for $\ell^2 > \kappa^2$, as
\begin{equation}\label{eq: K cent Coul}
\vec K = \mathcal{U}(-\psi) \left[ \sqrt{1 - \frac{\kappa^2}{\ell^2}} \vec p \times \vec \ell + \left(\frac{\ell^2 - \kappa^2 - \ell \sqrt{\ell^2 - \kappa^2}}{r} + \kappa E \right) \hat r \right ]
\end{equation}

For the critical value $\ell = \left|\kappa\right|$ for which the centrifugal barrier disappears, Eq.(\ref{eq: geneq Coul}) becomes
 \begin{equation} \label{eq: criteq Coul}
\left( \frac{1}{r^2} \frac{dr}{d\theta} \right)^2 + \frac{2E}{\kappa r} = \frac{E^2 - m^2}{\kappa^2}
 \end{equation}
with the solution
 \begin{equation} \label{eq: crit orbit Coul}
r = \frac{2\kappa E}{E^2 - m^2 - E^2 \theta^2}
 \end{equation}
The LRL vector~(\ref{eq: K cent Coul}) then reduces to
\begin{equation}\label{eq: K crit Coul}
\vec K = \mathcal{U}(-\theta) \kappa E \hat r
\end{equation}
In the anti-centrifugal case $\ell <\left| \kappa\right|$ the solution for the orbit becomes
 \begin{equation} \label{eq: nocent orbit Coul}
r = \frac{\kappa^2 - \ell^2}{E\kappa - K\cosh \left(\sqrt {\frac{\kappa^2}{\ell^2} - 1} \theta \right)}
 \end{equation}
Expressing $\theta$ in terms of $r$ and substituting in the generic LRL vector~(\ref{eq: vec K general 2}), will then yield the LRL vector in terms of $r$.

\vskip20pt

\subsection{\label{subsec: PN}The generalized Laplace-Runge-Lenz symmetry in post-Newtonian Kepler-Coulomb systems}

As a second example we derive in the following the generalized LRL vector for a 2-particle system in the post-Newtonian approximation either for electromagnetic (Darwin \cite{Darwin20,LLfields75}) or gravitational (Einstein-Infeld-Hoffman \cite{EIH38,LLfields75}) interactions. As for the former case, LRL vectors were considered in these systems only as rotating classical LRL vectors (see, \textit{e.g.}, \cite{BarkerO'Connell75}). The only time that constant LRL vectors were considered for these systems was, to the author's best knowledge, by Arg\"{u}eso and Sanz \cite{ArgSanz84}, who computed the LRL vector by quite cumbersome means. Our derivation, applying the methods developed in the preceding sections, is, to our belief, more elegant and illuminating. It is also used to show a different way of application.

Consider a 2-particle system with masses $m_1,m_2$ and possible electrical charges $e_1,e_2$. The post-Newtonian Hamiltonian for the two interactions in the CM system with internal polar canonical variables is given by \cite{LLfields75}
 \begin{equation} \label{eq: PN-EMG H}
H = \frac{p^2}{2\mu} - \frac{p^4}{8\nu^3 c^2} + \frac{\kappa}{r} +
\frac{\kappa}{2m_1 m_2 c^2 r} \left[
\left( 2 + \alpha \right) p_r^2 + \left( 1 + \alpha \right)
\frac{p_\theta^2}{r^2}
\right] + \frac{\alpha \kappa^2}{6 M_o c^2 r^2}
 \end{equation}
with $p^2 = p_r^2 + p_\theta^2/r^2$, $M_o = m_1 + m_2$,  $\mu = m_1 m_2 / M_o$, and
\[
\frac{1}{\nu^3} \equiv \frac{1}{m_1^3} + \frac{1}{m_2^3} = \frac{1}{\mu^3} - \frac{3}{M_o \mu^2}
\]
$\kappa = e_1 e_2$ and $\alpha = 0$ or $\kappa = -G m_1 m_2$ and $\alpha = 3M_o / \mu$ for electromagnetic or gravitational systems, respectively.

Instead of proceeding as before, namely computing $r_m$ from Eq.(\ref{eq: rm E el}) and then computing $K$ via Eq.(\ref{eq: K vec 1 gen}), we use a slightly alternative way. With constant $H = E'$ and $p_\theta = \ell$, $p_r^2$ is explicitly deduced from the hamiltonian (\ref{eq: PN-EMG H})
 \begin{equation} \label{eq: PN-EMG pr2}
p_r^2 \approx 2\mu E' - \frac{2\mu \kappa}{r} +
\left( 1 - \frac{3\mu}{M_o} \right) \frac{E'^2}{c^2} -
\left[ 1 - \frac{\left( 1 - \alpha \right) \mu}{M_o} \right] \frac{2\kappa E'}{c^2 r} +
\left[ 1 + \frac{\left( 3 + 5\alpha \right) \mu}{3M_o} \right] \frac{\kappa^2}{c^2 r^2} + \frac{\kappa \ell^2}{M_o c^2 r^3} - \frac{\ell^2}{r^2}
 \end{equation}
From Hamilton's equations
\begin{eqnarray}
\frac{dr}{dt} = \frac{\partial H}{\partial p_r} =
 \left[ \frac{1}{\mu} - \frac{p^2}{2 \nu^3 c^2} + \frac{\left( 2 + \alpha \right) \kappa }{m_1 m_2 c^2 r} \right] p_r  \nonumber \\
\frac{d\theta}{dt} = \frac{\partial H}{\partial p_\theta} =
\left[ \frac{1}{\mu} - \frac{p^2}{2\nu^3 c^2} + \frac{\left( 1 + \alpha \right) \kappa }{m_1 m_2 c^2r} \right] \frac{p_\theta}{r^2}
  \nonumber
\end{eqnarray}
it is easy to see that $p_r$ may be written as
 \begin{equation} \label{eq: pr - du/dtheta}
p_r = - \ell \frac{d u}{d\theta}
 \end{equation}
where, correct to terms up to order $1/c^2$,
 \begin{equation} \label{eq: PN-EMG u}
u = \frac{1}{r} - \frac{\kappa }{2 M_o c^2 r^2}
 \end{equation}
Substituting (\ref{eq: pr - du/dtheta}) and (\ref{eq: PN-EMG u}) in Eq.(\ref{eq: PN-EMG pr2}) and keeping terms up to order $1/c^2$ then yields the equation
 \begin{equation} \label{eq: PN-EMG pr-theta eq}
\left(1 - \delta\right)^{-2} \left( \frac{du}{d\theta} \right)^2 + \left( u - u_o \right)^2 = B^2
 \end{equation}
where
\begin{eqnarray}
&& u_o = - \left[ {1 + \left( {1 + \frac{{5\alpha \mu }}{{3{M_o}}}} \right)\frac{{{\kappa ^2}}}{{{\ell ^2}{c^2}}}} \right]\frac{{\mu \kappa }}{{{\ell ^2}}} - \left[ {1 - \frac{{\left( {1 - \alpha } \right)\mu }}{{{M_o}}}} \right]\frac{{\kappa E'}}{{{\ell ^2}{c^2}}}
 \nonumber \\
&& B^2 = \left\{ {1 + \left[ {2 + \frac{{\left( {8\alpha  - 3} \right)\mu }}{{3{M_o}}}} \right]\frac{{{\kappa ^2}}}{{{\ell ^2}{c^2}}}} \right\}\frac{{2\mu E'}}{{{\ell ^2}}} + \left( {1 - \frac{{3\mu }}{{{M_o}}}} \right)\frac{{E{'^2}}}{{{\ell ^2}{c^2}}} + \left[ {1 + \left( {1 + \frac{{5\alpha \mu }}{{3{M_o}}}} \right)\frac{{2{\kappa ^2}}}{{{\ell ^2}{c^2}}}} \right]{\left( {\frac{{\mu \kappa }}{{{\ell ^2}}}} \right)^2}  \nonumber \\
&& \delta = \left( {1 + \frac{{5\alpha \mu }}{{3{M_o}}}} \right)\frac{{{\kappa ^2}}}{{2{\ell ^2}{c^2}}}
 \nonumber
\end{eqnarray}
The solution, with the condition that $\theta = 0$ at a perihelion, is
 \begin{equation} \label{eq: PN-EMG u-theta}
u = u_o + B \cos \left[ \left(1 - \delta \right) \theta \right]
 \end{equation}
which immediately yields
 \begin{equation} \label{eq: PN-EMG pr-theta}
p_r =  - \ell \frac{du}{d\theta} = \left( 1 - \delta \right) \ell B\sin \left[ \left( 1 - \delta \right) \theta \right]
 \end{equation}
and the magnitude $K\left(E', \ell^2\right)$ [Eq.(\ref{eq: K def gen})]
 \begin{equation} \label{eq: PN-EMG K mag}
K = \ell {\left( {\frac{{d{p_r}}}{{d\theta }}} \right)_{r = {r_m}}} = {\left( {1 - \delta } \right)^2}{\ell ^2}B\cos {\left[ {\left( {1 - \delta } \right)\theta } \right]_{\theta  = 0}} = {\left( {1 - \delta } \right)^2}{\ell ^2}B
 \end{equation}

The constant LRL vector of the system is obtained simply by substituting its magnitude and expression (\ref{eq: PN-EMG pr2}) for $p_r$ in Eq.(\ref{eq: vec K general 2}),
\[
\vec K = \frac{{{{\left( {1 - \delta } \right)}^2}\ell B\sin \theta }}{{{p_r}}}\vec p \times \vec \ell  + {\left( {1 - \delta } \right)^2}{\ell ^2}B\left( {\frac{{\cos \theta }}{r} - \frac{{\ell \sin \theta }}{{{r^2}{p_r}}}} \right)\vec r
\]
To transform this expression to something of a more familiar form, similar to the classical one, it is convenient to introduce, as before, the virtual angle $\varphi = \left(1 - \delta\right) \theta$ so that $\left(r,\varphi\right)$ is the polar coordinate frame rotating with the orbit. The LRL vector which is fixed in the rotating frame is obtained by replacing $\theta$ with $\varphi$,
\[
\vec K' = \frac{{{{\left( {1 - \delta } \right)}^2}\ell B\sin \varphi }}{{{p_r}}}\vec p \times \vec \ell  + {\left( {1 - \delta } \right)^2}{\ell ^2}B\left( {\frac{{\cos \varphi }}{r} - \frac{{\ell \sin \varphi }}{{{r^2}{p_r}}}} \right)\vec r
\]
which yields, using Eqs.~(\ref{eq: PN-EMG u}) and (\ref{eq: PN-EMG pr-theta}),
\[
\vec K' = \left( {1 - \delta } \right)\vec p \times \vec \ell  + \left\{ \left[ 1 + \left( \frac{1}{\mu} - \frac{1 - \alpha}{M_o} \right) \frac{E'}{c^2} \right] \frac{\mu \kappa}{r} - \left( 1 + \frac{5\alpha \mu}{3 M_o} \right)
\frac{\kappa^2}{2 c^2 r^2} - \frac{\kappa \ell^2}{2M_oc^2r^3} \right\} \vec r
\]
Using the identity ${\mathcal U}\left( \varphi \right)\vec K' = {\mathcal U}\left( \theta  \right)\vec K = K\hat r$ which follows from Eq.(\ref{eq: vec K general 1}), the constant LRL vector $\vec K$ is then obtained by rotating $\vec K'$ back to the fixed system,
\begin{eqnarray} \label{eq: PN-EMG K vec}
\vec K && = \mathcal{U}\left( - \theta \right)\mathcal{U}\left( \varphi  \right) \vec K' = \mathcal{U}\left( \varphi - \theta \right) \vec K' = \mathcal{U}\left( - \delta \theta \right) \vec K' \approx \left( 1 - \delta \theta \hat \ell \times \right) \vec K' =
 \nonumber \\
&&  = \left( 1 - \delta \theta \hat \ell \times \right) \cdot
\left\{
\left( {1 - \delta } \right)\vec p \times \vec \ell  +
\left\{ \left[ 1 + \left( \frac{1}{\mu} - \frac{1 - \alpha}{M_o} \right) \frac{E'}{c^2} \right] \frac{\mu \kappa}{r} - \left( 1 + \frac{5\alpha \mu}{3 M_o} \right)
\frac{\kappa^2}{2 c^2 r^2} - \frac{\kappa \ell^2}{2M_oc^2r^3} \right\} \vec r
\right\}
\end{eqnarray}

Finally, the self PB of $\vec K$ are very simply computed by substituting the squared magnitude
\[
{\vec K^2} = {\left( {1 - \delta } \right)^4}{\ell ^4}{B^2} \approx \left[ {{\ell ^2} - \frac{{\left( {2\alpha  + 3} \right)\mu {\kappa ^2}}}{{3{M_o}{c^2}}}} \right]2\mu E' + \left( {1 - \frac{{3\mu }}{{{M_o}}}} \right)\frac{{E{'^2}{\ell ^2}}}{{{c^2}}} + {\mu ^2}{\kappa ^2}
\]
in Eq.(\ref{eq: PB KK}) to obtain
 \begin{equation} \label{eq: PN-EMG K PB}
\left\{ {{K^i},{K^j}} \right\} =  - \frac{{\partial \left( {{{\vec K}^2}} \right)}}{{\partial \left( {{\ell ^2}} \right)}}{\varepsilon ^{ijk}}{\ell ^k} =  - \left[ {2\mu E' + \left( {1 - \frac{{3\mu }}{{{M_o}}}} \right)\frac{{E{'^2}}}{{{c^2}}}} \right]{\varepsilon ^{ijk}}{\ell ^k} \, ,
 \end{equation}
thus avoiding much tedious work that would be required for the computation of the PB  directly from the explicit expression (\ref{eq: PN-EMG K vec}). It is interesting to note that the rhs of Eq.(\ref{eq: PN-EMG K PB}) depends only on the energy, and is independent of any other detail of the interaction. As well, it is easy to verify that the results of both Subsections coincide for the post-Newtonian approximation of a Coulomb system (one charge with infinite mass).

\vskip30pt

\section{\label{sec: CMPN}The relativistic origin of LRL symmetry in general Newtonian centrally-symmetric systems}

The last aspect of the LRL symmetry in general rotationally symmetric systems to be discussed here is its relativistic origin. After having been shown by Dahl\cite{Dahl97} that the classical Newtonian LRL vector (\ref{eq: KC LRL}) emerges naturally from the computation of the Lorentz boost in post-Newtonian extensions of the Kepler-Coulomb interaction, we show in the following that the LRL vector (\ref{eq: K vec 2}) of general Newtonian centrally-symmetric systems emerges, in a similarly natural way, from the Lorentz boost in the corresponding post-Newtonian extensions. Consequently, Dahl's original result is not particular to classical Kepler-Coulomb systems only, but (at least) to all centrally symmetric ones.

It should be emphasized that unlike Sec.\ref{subsec: PN}, where we looked for the post-Newtonian LRL vector, here we look for the Newtonian LRL vector which is derived from the post-Newtonian Lorentz boost, in a completely different procedure.

We start by reviewing the post-Newtonian extensions of Newtonian
systems with general central interactions \cite{Gaida82}.
Consider a 2-particle system with masses $m_1,m_2$, spatial
coordinates $\vec x_1 , \vec x_2$, linear momenta $\vec p_1 , \vec
p_2$ and Newtonian central potential $U_o(r)$ with $\vec r = \vec
x_1  - \vec x_2$. In post-Newtonian extensions the total linear
and angular momenta maintain their Newtonian form,
\[
\vec P = \sum\limits_a {\vec p_a } = \vec p_1 + \vec p_2 \qquad ,
\qquad  \vec J = \sum\limits_a {\vec x_a \times \vec p_a} \quad ,
\]
while a scalar interaction $U_1 \left( {\vec r , \vec p_a }
\right)$ of order $1/c^2$ is added to the total energy together
with the kinetic terms of the same order,
 \begin{equation} \label{eq: PN energy}
P^o = E = \sum\limits_a \left(m_a c^2 + \frac{p_a^2}{2m_a} - \frac{p_a^4}{8m_a^3 c^2} \right) + U_o \left( r \right) + \frac{1}{c^2} U_1 \left( \vec r , \vec p_a \right)
 \end{equation}
The most general form for the Lorentz boost in the post-Newtonian extension is
 \begin{equation} \label{eq: PN boost}
 \vec N = \sum\limits_a {\left[ m_a + \frac{p_a^2}{2m_a
c^2} + \frac{1}{2c^2} U_o \left( r \right) \right] \vec x_a} + \frac{1}{c^2} \vec \Psi - \vec P t
\end{equation}
with $\vec \Psi \left( \vec r , \vec p_a \right)$ another unknown
vector interaction term.

The properties of the unknowns -- $U_1$ and $\vec \Psi$ -- are
determined by the requirement that $\vec P$, $P^o$, $\vec J$ and
$\vec N$ satisfy, to order $1/c^2$, the Lorentz-Poincar\'{e}
Lie-Poisson brackets. In particular, the PB $\left\{ \vec N , P^o
\right\} = \vec P$ imply
 \begin{equation}\label{eq: dN/dt}
\sum\limits_a \frac{\partial U_1}{\partial \vec v_a} + \left\{
\vec \Psi , H_{\rm N} \right\} = \frac{1}{2r} U_o'\left( r \right) \left[
{\vec r \cdot \left( \vec v_1 + \vec v_2 \right)} \right] \vec r -
\frac{1}{2}U_o \left( r \right) \left( \vec v_1 + \vec v_2 \right)
\end{equation}
with $H_{\rm N}$ being the Newtonian Hamiltonian (\ref{eq: hamiltonian 2-body}) with potential $U_o \left( r \right)$. We need not go further into the details of $\vec \Psi$ and $\tilde U_1 \left(\vec r , \vec v\right)$, because Eq.(\ref{eq: dN/dt}) is all
that is required for the following.

To obtain the LRL vector, we start from the fact that the Newtonian centre-of-mass
 \begin{equation} \label{eq: Newton CM}
 \vec X_{\rm N} =  \frac{m_1 \vec x_1 + m_2 \vec x_2}{M_o}
 \end{equation}
($M_o = m_1 + m_2$) is constant only in the non-relativistic limit, and look for an internal vector $\vec R$ which satisfies, in the centre-of-mass (CM) frame,
\begin{equation} \label{eq: R eqn PN 1}
\frac{d\vec R}{dt} = \frac{d\vec X_{\rm N}}{dt} = \frac{m_1 \vec v_1  + m_2 \vec v_2}{M_o}
\end{equation}
The CM reference frame is defined by $\vec P = \vec p_1 + \vec
p_2$ = 0, without fixing the origin. Substituting the  post-Newtonian velocities
 \begin{equation} \label{eq: PN va}
 \vec v_a = \frac{\partial P^o}{\partial \vec p_a} = \left( 1 - \frac{p_a^2}{2 m_a^2 c^2}
\right) \frac{\vec p_a}{m_a}
 + \frac{1}{c^2} \frac{\partial U_1}{\partial \vec p_a} \, ,
 \end{equation}
with the the CM condition $\vec p_1 = - \vec p_2 = \vec p$ and Eq.(\ref{eq: dN/dt}), Eq. (\ref{eq: R eqn PN 1}) becomes
\begin{eqnarray} \label{eq: R eqn PN 2}
\frac{d\vec R}{dt} & = & \frac{\left( m_1 - m_2 \right) p^2}{2m_1^2 m_2^2 c^2}
\vec p + \frac{1}{M_o c^2} \sum\limits_a
\frac{\partial U_1}{\partial \vec v_a} = \nonumber \\
 & = & \frac{\left( m_1 - m_2 \right) p^2}{2m_1^2 m_2^2
c^2} \vec p + \frac{1}{M_o c^2} \left\{ \frac{1}{2r} U_o'\left(
r \right) \left[ \vec r \cdot \left( \vec v_1 + \vec v_2 \right)
\right] \vec r - \frac{1}{2} U_o \left( r \right) \left( \vec v_1
+ \vec v_2 \right) \right\} - \frac{1}{M_o c^2} \left\{ \vec \Psi , H_{\rm N} \right\} = \nonumber \\
  & = & \frac{m_1 - m_2}{2m_1 m_2 M_o c^2} \left\{
\left[ \frac{p^2}{\mu} + U_o \left( r \right) \right] \vec p -
\frac{1}{r} U_o' \left( r \right) \left( \vec r \cdot \vec p
\right)\vec r
\right\} - \frac{1}{M_o c^2} \frac{d\vec \Psi}{dt}
\end{eqnarray}
with $\mu = m_1 m_2 / M_o$ the Newtonian reduced mass.

Since the rhs is already of order $1/c^2$, we look for solutions of this equation in which $\vec R$ is an internal vector of order $1/c^2$. Newtonian relations
and equations of motion are then sufficient in the following. As in Dahl's result, Eq.(\ref{eq: R eqn PN 2}) is now expected to have two independent solutions, with the difference between them being proportional to the LRL vector. One solution is obtained from $\vec X_{\rm N}$ and the Lorentz boost $\vec N$. Expressing the particles' coordinates in a way similar to the Newtonian relations
 \begin{equation} \label{eq: coor from Ro}
\vec x_1 = \vec X_{\rm N} + \frac{m_2}{M_o} \vec r \qquad  ,  \qquad
\vec x_2 = \vec X_{\rm N} - \frac{m_1}{M_o} \vec r \quad ,
 \end{equation}
the post-Newtonian Lorentz boost (\ref{eq: PN boost}) becomes, in
the CM frame,
 \begin{eqnarray} \label{eq: PN boost CM}
\vec N && = \sum\limits_a {\left[ m_a + \frac{p^2}{2m_a c^2} + \frac{1}{2c^2} U_o \left( r \right) \right] \vec x_a} + \frac{1}{c^2}
\vec \Psi \left( \vec r , \vec p \right) =
\nonumber  \\
&& = M \vec X_{\rm N} + \frac{m_2 - m_1}{2\mu M_o c^2} \left[ p^2 +
\mu U_o \left( r \right) \right] \vec r + \frac{1}{c^2} \vec
\Psi \left( \vec r , \vec p \right)
 \end{eqnarray}
where
 \begin{equation} \label{eq: M PN}
M = M_o + \frac{p^2}{2\mu c^2} + \frac{1}{c^2} U_o \left( r \right)
 \end{equation}
is the total relativistic mass. Then the first solution in the post-Newtonian approximation is simply
 \begin{equation} \label{eq: PN R1}
\vec R_1 = \vec X_{\rm N} - \frac{\vec N}{M} = \frac{m_1 - m_2}{2\mu M_o^2 c^2} \left[ p^2 + \mu U_o \left( r \right) \right] \vec r - \frac{1}{M_o c^2} \vec
\Psi \left( \vec r , \vec p \right)
 \end{equation}

To obtain a second, independent solution for Eq.(\ref{eq: R eqn PN 2}), we first employ the Newtonian equations of motion and write the equation in the form
\begin{eqnarray} \label{eq: R eqn PN 4}
 \frac{d\vec R}{dt} && = \frac{m_1 - m_2}{2\mu M_o^2 c^2} \left\{ \frac{d}{dt} \left[ \left( \vec r \cdot \vec p \right) \vec p \right]
+ \left[ U_o \left( r \right) + rU_o' \left( r \right) \right]
\vec p \right\} - \frac{1}{M_o c^2} \frac{d\vec \Psi}{dt} =
\nonumber  \\
  && = \frac{m_1 - m_2}{2\mu M_o^2 c^2} \left\{ \frac{d}{dt} \left[ \left( \vec r \cdot \vec p \right) \vec p + \mu
\left(rU_o\right)' \vec r \right] - p_r \left[ rU_o \left( r
\right) \right]'' \vec r \right\} - \frac{1}{M_o c^2} \frac{d\vec \Psi}{dt}
\end{eqnarray}
Then, with the vector $\vec W \left(\vec r , \vec p \right)$ satisfying Eq.(\ref{eq: dW/dt}), the second solution $\vec R_2$ is identified as
\begin{equation} \label{eq: Ro int}
\vec R_2 = \frac{m_1 - m_2}{2\mu M_o^2 c^2} \left[
\left( \vec r \cdot \vec p \right) \vec p + \mu
\left(rU_o\right)' \vec r - \vec W \right] - \frac{1}{M_o c^2}
\vec \Psi
\end{equation}

The difference between the two solutions,
 \begin{equation} \label{eq: R1-R2 PN}
\vec R_1 - \vec R_2 = \frac{m_1 - m_2}{2 \mu M_o^2 c^2} \left[ \vec p
\times \vec \ell - \mu rU_o' \left( r \right) \vec r + \vec W
\right] \, ,
 \end{equation}
is clearly recognized as being proportional to the LRL vector of the corresponding Newtonian system (\ref{eq: K vec 2}). We notice that it is independent of the post-Newtonian corrections $U_1$ and $\vec \Psi$, depending only on the Newtonian
limit, so it is the same for all possible post-Newtonian extensions of the same Newtonian potential. This result verifies the relativistic origin of the LRL symmetry for all Newtonian centrally-symmetric 2-body systems.

An extra benefit of Dahl's procedure is fixing the preferred direction of the LRL vector. Although the classical LRL vector (\ref{eq: KC LRL}), as historically constructed and used over the years, is directed towards the perihelion, there is nothing in the LRL symmetry by itself, as was already discussed above (Sec.\ref{sec: algeprop}), that distinguishes any particular direction for it. Choosing the direction of the LRL vector in the explicit construction in Sec.\ref{sec: expcon} towards a closest approach was based more on aesthetical reasons (regularity of the coefficients) rather than on more profound ones. It is really the relativistic consideration, even at the post-Newtonian level, that distinguishes this direction as the preferred one, because this is the direction of the vector that appears in Eq.(\ref{eq: R1-R2 PN}). It may be argued, of course, that as an integral, the vector $\vec R_2$ is anyway defined up to an arbitrary addition, so it may result in any desired direction. Still, this is an arbitrary addition, while without it the natural direction that appears is towards a (generalized) perihelion.

\vskip30pt

\section{\label{sec: discussion}Discussion and concluding remarks}

In the present paper we brought and discussed, from several angles, new evidence that support  the generality and universality of the Laplace-Runge-Lenz symmetry.

The main aspects that were discussed are :

\begin{enumerate}
\item{
The independence of the symmetry upon the type of interaction, requiring only internal rotational symmetry. This was verified by the propositions of Sec.\ref{sec: genLRL}, leading to Eq.(\ref{eq: PB KK}) and culminating in Eq.(\ref{eq: PBAA}), demonstrating that the symmetry is always $o(4)$ or $o(3,1)$ depending only on the energy state (bound or unbound systems). The dependence of the symmetry only upon the energy state is emphasized by the results of Sec.\ref{sec: Coul} for different types of relativistic systems, where the PB (\ref{eq: PB KK Coul}) and (\ref{eq: PN-EMG K PB}) are independent of other details of the orbit (like the status of the potential barrier in Sec.~\ref{subsec: Coul}), except the energy.

These two propositions contain the fundamental ingredients of the internal symmetry, incorporating the core property for the generalization of the LRL symmetry to arbitrary rotationally symmetric systems. Eq.(\ref{eq: PB KK}) thus becomes the fundamental equation of the LRL symmetry.

The generality of the symmetry is strengthened by the fact that the proof of the propositions uses only generic properties of the Poisson brackets, independent of any particular recipe for their computation.
}
\item{
Definition of the LRL vector -- the generator of the LRL symmetry -- as a constant vector (not only piece-wise conserved) even if the orbit is open, for all centrally-symmetric 2-body systems. As an application, the LRL vector in relativistic Coulomb systems and post-Newtonian electromagnetic or gravitational systems was computed.
}
\item{
Demonstration of the relativistic origin of the LRL vector in general rotationally-symmetric systems. This is a generalization of Dahl's result, which referred only to Newtonian $1/r$ potentials, into all (arbitrary) centrally symmetric potentials.
}
\end{enumerate}

Let us discuss now the picture that these aspects combine together.

The relativistic origin of the LRL symmetry leads us to focus with special interest at its appearance in relativistic systems. Relativistic systems, even simpler ones like the relativistic Coulomb system or the post-Newtonian systems discussed in Sec.\ref{sec: Coul}, are characteristically endowed with bound states with open orbits. Still, the LRL vector that emerges from Dahl's procedure is constant. Thus it was important to insure that we are equipped with valid definition for constant LRL vectors, as was done in Sec.\ref{sec: expcon}.

It has also been shown that although the LRL symmetry, by itself, does not imply any particular direction for the LRL vector, the extension to relativistic systems yields a preferred direction -- towards the perihelia of the orbits, as in the classical Kepler-Coulomb case. This follows, for centrally symmetric systems, from the relation between the post-Newtonian vector (\ref{eq: R1-R2 PN}) and the Newtonian LRL vector (\ref{eq: K vec 2}).

The LRL symmetry is an internal symmetry -- it affects the relative state of the particles in a system, but it does not affect the system as a whole (the global, CM motion, remains unaffected). In Dahl's procedure (both in its original form\cite{Dahl97} and in the present generalization), the LRL vector is derived from the Lorentz boost. This fact points to the possibility that the LRL symmetry is internally associated with global Lorentz transformations, analogously to the way that internal rotations are associated with global rotations. In other words, this suggests that the LRL and rotational symmetries are attached together \textit{internally} in the same way that global rotations and Lorentz transformations form together the generalized rotations in Minkowski space-time (see the diagram below). This also explains why the internal symmetry generated by the internal rotations and the LRL vector is $o(4)$ or $o(3,1)$ :
\[
\begin{array}{cccc}
 \textrm{Global space-time symmetry :} \quad &
\begin{array}{c}
\textrm{global} \\ \textrm{rotations}
\end{array}
 & \, + \, &
 \begin{array}{c}
\textrm{Lorentz} \\ \textrm{transformations}
\end{array}
 \\
  &  &  &  \\
  & \updownarrow &  & \updownarrow  \\
  &  &  &  \\
 \textrm{Internal symmetry :} \quad &
\begin{array}{c}
\textrm{internal} \\ \textrm{rotations}
\end{array}
 &  \, + \, &
 \textrm{LRL}
\end{array}
\]
Yet another way to look at the association between the LRL symmetry and the Lorentz transformations is by noting that the transformations generated by the LRL vector change, for a given value of the total energy $E$, the internal angular momentum and thus the internal configuration of the system, in excellent analogy with the Lorentz transformations changing globally the way the system moves as a whole.

The LRL symmetry is therefore found to be an integral part of the internal relativistic symmetry. A detailed discussion of the r\^{o}le that it plays in the internal symmetry of Lorentz-Poincar\'{e} symmetric systems and its implications on the relativistic centre-of-mass is given elsewhere \cite{InternalSym}.

Finally, it should be noted that although the LRL symmetry is known so far to be found only in 2-body systems, it follows from Sections \ref{sec: genLRL} and \ref{sec: algeprop} that all that is required for its existence is rotational symmetry and the existence of PB. Then, following the preceding discussion, the LRL symmetry is an integral part of the internal symmetry of the system. For 2-body systems, the symmetry generated by $\vec\ell$ and the LRL vector is the full internal symmetry. For larger, $N(\ge 3)$-body systems, the internal symmetry may be larger, but the LRL symmetry is expected to be an integral part of it. Therefore, at least in principle, the LRL symmetry may well apply also to larger ($N \ge 3$) systems, as long as they are endowed with rotational symmetry. Indeed, two particular cases of the LRL symmetry in many body systems have already been discussed elsewhere \cite{InternalSym,manybodyPN}.

\vskip20pt

\rule{10cm}{1pt}



\begin{thebibliography}{10}
\bibitem{Bacry91}
As pointed out by Bacry (H. Bacry, "Comment on "The relationship between the symmetries of and the existence of conserved vectors for the equation $\ddot{r} + f(r)l + g(r) = o$"", J.Phys. A: Math. Gen. \textbf{24}, 1157 (1991)), in the case of 2-body Newtonian systems with central potential $U(r)$, $\ell$ is maximal in circular motion only if $r^3 U'(r)$ is an increasing function of $r$. But this is also the condition for the stability of the circular orbit, so we proceed assuming that circular orbits are stable. The opposite case was not studied.
\bibitem{Fradkin fn}
Fradkin (Ref.~\onlinecite{Fradkin67}) arrived at an expression similar to Eq.(\ref{eq: Casimir1}), but he regarded the Casimir invariant $\mathcal{C}_1 (E)$, which he denoted $P^2(E)$, as arbitrary, which evidently is not the case.
\end{thebibliography}
    \end{document}